\documentclass[aps,pre,amsmath,superscriptaddress,amssymb,reprint,floatfix]{revtex4-1}
\usepackage{amsmath}
\usepackage{amsfonts}%
\usepackage{amssymb}
\usepackage{epsfig}
\usepackage{color}
\usepackage{graphics}
\usepackage{graphicx}
\usepackage{ifpdf} 
\usepackage{epsfig}
\usepackage{multirow}
\usepackage[outdir=./]{epstopdf}
\usepackage{subfigure}  
\usepackage[colorlinks=true,linkcolor=red]{hyperref}   
\usepackage{bm}
\usepackage[colorinlistoftodos]{todonotes}
\usepackage{lipsum}
\usepackage[utf8]{inputenc}

\begin{document}
	
	\title{Effective interactions between inclusions in a chiral active bath inside a channel}
	
	\author{Abdolhalim Torrik}
	\email{yashar.torrik@gmail.com}
	\affiliation{Department of Physical and Computational Chemistry, Shahid Beheshti University, Tehran 19839-9411, Iran} 
	
	\author{Ali Naji}
	\email{a.naji@ipm.ir}
	\affiliation{School of Physics, Institute for Research in Fundamental Sciences (IPM), Tehran 19395-5531, Iran}	
	\affiliation{School of Nano Science, Institute for Research in Fundamental Sciences (IPM), Tehran 19395-5531, Iran}

	\author{Mahdi Zarif}
	\email{m\_zarif@sbu.ac.ir (corresponding author)}
	\affiliation{Department of Physical and Computational Chemistry, Shahid Beheshti University, Tehran 19839-9411, Iran}
	
	\date{\today}
	
	\begin{abstract}
		Colloidal inclusions suspended in a bath of smaller particles experience an effective bath-mediated attraction at small intersurface separations, which is known as the depletion interaction. In an active bath of nonchiral self-propelled particles, the effective force changes from attraction to repulsion; an effect that is suppressed, when the active bath particles are chiral. Using Brownian Dynamics simulations, we study the effects of channel confinement and bath chirality on the effective forces and torques that are mediated between two inclusions that may be fixed within the channel or may be allowed to rotate freely as a rigid dimer around its center of mass.  We show that the confinement has a strong effect on the effective interactions, depending on the orientation of the dimer relative to the channel walls. The active particle chirality leads to a force imbalance and, hence, a net torque on the inclusion dimer, which we investigate as a function of the bath chirality strength and the channel height. 
	\end{abstract}

	
	\maketitle
	
	\section{Introduction}
	\label{sec:Intro}
	
Active particles feature internal mechanisms that enable them to take up ambient free energy and execute self-propelled motion. Being thus driven to a state of nonequilibrium, a collection of such particles can exhibit intriguing phenomena, such as self-organized assemblies and collective motions \cite{likos2001effective,romanczuk2012active,bechinger2016active,lekkerkerker2011depletion,
shields2017evolution,khadka2018active,shklyaev2017convective,hagan2016emergent,Marchetti:RMP2013,ramaswamy:ARCMP2010,lowen2016chirality,breier2016spontaneous,shaebani2020computational}. In biology, most microorganisms have evolved to perform active motion to explore their (often aqueous) surroundings \cite{Hirokawa:NRMCB2009,Woolley:Reproduction2003,SHACK:BMB1974}. Recently, a wide range of artificial nano/microswimmers and active particles, such as Janus colloids, have also been manufactured, utilizing different mechanisms for self-propelled motion to facilitate advanced applications, such as cargo and drug delivery \cite{Walther:CR2013,Jiang:AM2010,Zhang:langmuir2017,bunea2020recent,tsang2020roads,rao2015force,zheng2017orthogonal}.
	
	When active particles are of asymmetric shape, the direction of self-propulsive force (motion) may deviate from the axis of geometry, leading to an induced torque on the particles \cite{Friedrich:PRL2009,Kummel:PRL2013,Lowen:EPJST2016}. Torques, can also be applied to swimmers externally, for example, using magnetic fields \cite{Baraban:ANano2013}. Active particles may also carry internal mechanisms of torque generation, giving them a persistent rotary motion. Such chiral self-propellers can thus exhibit circular (helical) trajectories in two (three) dimensions \cite{Breier:PRE2014,Li:PRE2014,Mijalkov:SoftMatter2015,Crespi:PRL2015,Ai:SR2016}.
	 
	Many microorganisms such as sperm cells in oviducts and \emph{Escherichia coli} in guts swim in confined geometries. Nonequilibrium surface accumulation of active particles have thus received mounting interest as a peculiar and yet generic effect realized in active systems \cite{Elgeti:EPJST2016}. This effect has been observed experimentally for \emph{E. coli} between two glass plates \cite{Berke:PRL2008} and numerically for cells with a rigid flagellar filament, indicating the dependence of accumulation on activity (self-propulsion speed) and cell size \cite{ROTHSCHILD:Nat1963,Li:PRE2011}. 
	Interactions between active systems and confining geometries can cause interesting collective phenomena, such as vertical collective motion of the bacterial suspension \cite{Elgeti:EPJST2016}.
	Understanding the effects of confinement is important for application purposes, such as separation, trapping, and sorting of active particles \cite{Elgeti:EPJST2016}.
	
	In a bath of small colloids, encompassing larger colloidal inclusions subject to repulsive steric interactions, there are depletion zones around the larger inclusions from which the smaller particles are excluded. When the inclusions approach each other and their intervening gap is smaller than the diameter of smaller particles, the respective depletion zones start to overlap, and there will be an imbalance in the osmotic pressure acting on the opposing sides of each of the inclusions; hence, giving an effective attraction, known as the depletion interaction, between the inclusions \cite{Likos:PR2001,Lekkerkerker:book2011}. 

	There have been numerous studies regarding depletion-type interactions and their effects on phase behavior, self-assembly, and stability of colloidal suspensions \cite{Likos:PR2001,Lekkerkerker:book2011,Stradner:Nat2004,Savage:Science2006,Koumakis:SM2011}. It is important in many aspects from technological to medicinal and biological fields because of the emergent effects like coagulation, flocculation, and agglomeration of colloids. Many properties of depletants like shape, size, charge, and concentration can modify the depletion interaction. Several types of depletants have also been studied including spherical particles, disk-like particles, rod-like particles, polymers, and micelles to name a few \cite{Richetti:PRL1992,McNamee:Lng2004,Helden:Lang2004}. Properties of colloids can also play an important role in depletion interactions e.g. lock and key colloids or introduction of surface roughness which can reduce this type of interaction between them. Some experimental techniques such as atomic force microscopy, total internal reflection microscopy, surface force apparatus, optical tweezers, and neutron scattering have been applied to measure depletion forces (see Ref. \cite{XING:COCIS2015} as a short review and reference therein).
	
	
	In the case of colloidal particles suspended in an active-particle bath, apart from the steric effects previously mentioned, nonequilibrium dynamics of particles also have a role in depletion forces. In its nonequilibrium realization, depletion forces violate Newton's third law \cite{dzubiella2003depletion,hayashi2006law}; are anisotropic and can generate attractive or repulsive interactions. This interaction can be dependent on many variables e.g. self-propulsion magnitude, the distance between suspended colloids, the size ratio of colloids and active particles, shape of colloids and particles, confinements, geometrical obstacles, or chirality of active particles \cite{Likos:PRL2003,Angelani:PRL2011,Harder:JCP2014,Cacciuto:PRE2016,Cacciuto:SoftMatter2016,Leite:PRE2016,Steimel:PNAS2016,Zaeifi:JCP2017,Dolai:SoftMatter2018,Yang:JPCM2018,Liu:PRL2020,Feng:PRE2021,sebtosheikh2020effective,sebtosheikh2021noncentral}.
	
	
	In contrast to the case of passive (equilibrium) depletion interactions, the inclusion of large nonactive particles into a medium containing small, nonchiral, active particles leads to a repulsive depletion interaction at short distances between inclusions, followed by a weak repulsive hump at moderate separations (around two to three small particle diameters), and finally a decaying oscillatory tail at larger separations \cite{Harder:JCP2014,Zaeifi:JCP2017}. 
	The repulsion forces in the steady state are due to the accumulation of active particles between the inclusions, where they get trapped and a concentration gradient builds up, which leads to the reported repulsive force \cite{Harder:JCP2014,Zaeifi:JCP2017}. The said oscillating behavior in the effective interaction is connected with the sequential overlaps of the active-particles layers (rings) that form around the inclusions. The number of layers varies according to the bath activity, concentration of active particles, and the presence of any other boundaries. Also, the amplitude and range of the ring overlaps drastically exceed their equilibrium counterparts, as the persistent motion of active particles results in prolonged near-surface detention times and more extensive steric layering around the inclusions \cite{Elgeti:EPJST2016}. 
	
	Bath-mediated forces on the inclusions change as chirality is introduced to the bath particles. In general, in the case of chiral active particles, the number and density of particle rings are suppressed and the intervening region between the inclusions becomes less populated, leading to suppression of the repulsive force between them. As the chirality strength is increased (i.e., the circular trajectories traversed by the chiral active particles display smaller radii \cite{Xue:EPL2015,Ao:EPJST2014}), the effective interaction turns to a relatively long-ranged, chirality-induced attraction \cite{Zaeifi:JCP2017}. 
	
	The effect of confinement on the interaction between two colloidal inclusions in a bath containing active particles has been discussed in Ref. \cite{Zarif:PRE2020} by two of the present authors. Based on these findings, in addition to the layered structures around the inclusions, active bath particles are also attracted by the channel walls, where they form flat layers. The number of wall layers increases with the area fraction of active particles and at the expense of the active-particle rings around the inclusions, indicating longer persistence times of active particles along flat boundaries as compared with the convex boundaries of the inclusions. The wall-induced effects are found to be more prominent, when the inclusion dimer is placed in perpendicular configuration with respect to the channel walls as opposed to a parallel configuration. 

	In this work, we extend the previous work \cite{Zarif:PRE2020} by considering (1) a chiral active bath and (2) both the cases of a fixed and a freely rotating inclusion dimer in the bath, with all particle being confined with a planar channel. In the case of a freely rotating inclusion dimer, we model the dimer as a dumbbell, pinned from the middle to the center of the channel. Using Brownian dynamics simulations, we investigate the confinement-induced effects on the effective interaction force and torque acting on the inclusions. 
	
	The paper is organized as follows: In Section~\ref{sec:model-methods}, we introduce our model and methods, and in Section~\ref{sec:results}, we discuss the results for fixed (Section~\ref{sec:fixed}) and rotating (Section~\ref{sec:mobile}) inclusions in the chiral active bath. The paper is concluded in Section~\ref{sec:Conclusion}.
	
	\section{Model and methods}
	\label{sec:model-methods}
	
\begin{figure}[t!]
		\begin{center}
			\includegraphics[width=0.5\textwidth]{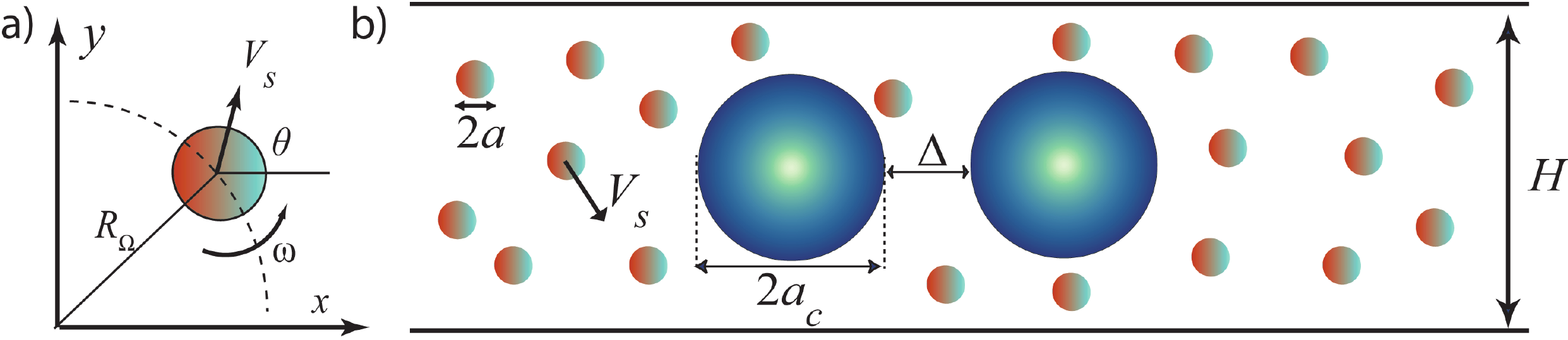}
			\caption{(a) Schematic view of a chiral active particle with self-propulsion
				speed $V_s$ and counterclockwise angular velocity $\omega$. (b) Two nonactive colloidal inclusions are placed at fixed positions with intersurface distance $\Delta$ in a bath of chiral (or nonchiral) active particles moving according to Eqs. \eqref{Eq:langevin_a},\eqref{Eq:langevin_b}.}
			\label{fig:Sketch}
		\end{center}
	\end{figure}

	Our model comprises two colloidal inclusions of radius $a_c$ and fixed center-to-center distance $\Delta$ inside a planar channel of height $H$ that also contains active Brownian particles of radius $a<a_c$ at area fraction $\phi$. Figure \ref{fig:Sketch}b shows the parallel configuration in which the fixed inclusion dimer in placed along the channel centerline. Other (fixed or mobile) configurations of interest for the dimer will be specified later. The system is studied in a two-dimensional setting, as widely used in the literature, given that most of the key features of active models can be reproduced efficiently in two dimensions \cite{Lauga:RPP009,Marchetti:RMP2013,Yeomans:EPJ2014,Elgeti:RPP2015,Lauga:ARFM2016,golestanian:SoftMatter2011, ramaswamy:ARCMP2010,cates:ARCMP2015,Romanczuk:EPJ2012,Zottl:JPCM2016,bechinger:RMP2016}.

	Active particles self-propel at constant speed $V_s$. At any given time $t$ their configuration is described by the position vectors ${\mathbf r}_i(t) = (x_i(t), y_i(t))$ and the orientation angles $\theta_i(t)$, from which their preferred direction of motion ${\mathbf n}_i=(\cos \theta_i, \sin \theta_i)$ is obtained (Fig. \ref{fig:Sketch}a). Particle configurations evolve in time according to the over-damped Langevin equations (see, e.g., Refs. \cite{Marchetti:RMP2013,Elgeti:RPP2015,Romanczuk:EPJ2012,Volpe:AJP2014})
	\begin{eqnarray}
		\dot{{\mathbf r}}_i &= &V_s{\mathbf n}_i-\mu_T\frac{\partial  U}{\partial {{\mathbf r}_i}}+\sqrt{2 D_T}\, {\boldsymbol \eta}_i(t),
		\label{Eq:langevin_a}
		\\
		\dot{\theta}_i&=&\omega + \sqrt{2D_R}\, \zeta_i(t),
		\label{Eq:langevin_b}
	\end{eqnarray}
	Here, $\mu_T$ is the translational mobility of active particles, $\mathbf{f_i}\equiv -{ \partial U}/{\partial {{\mathbf r}_i}}$ is the force acting on the $i$th particle due to interactions with other particles and the walls.
	$D_T$ is the (bare) translational coefficient of active particles fulfilling the Einstein-Smoluchowski-Sutherland relation $D_T = \mu_T k_{\mathrm{B}}T$ ($k_{\mathrm{B}}$ is the Boltzmann constant and $T$ is the ambient temperature). $D_R$ is the rotational diffusion constant of active particles satisfying the relation $D_R = 3D_T/\sigma^2$ in low Reynold's regime. $\omega$ is the intrinsic angular velocity of active particles with positive (negative) values corresponding to counterclockwise (clockwise) particle rotations. ${\boldsymbol \eta}_i(t)$ and ${\boldsymbol \zeta}_i(t)$ are independent, white, Gaussian translational and rotational noises with zero mean, $\langle {\eta}_i^\alpha(t) \rangle = \langle \zeta_i(t) \rangle=0$, and two-point correlations $ \langle {\eta}_i^\alpha(t) {\eta}_j^\beta(t') \rangle=_{ij}\delta_{\alpha\beta}\delta(t-t')$ and $ \langle \zeta_i(t) \zeta_j(t') \rangle=\delta(t-t')$, with $i, j$ denoting the active particle labels and $\alpha, \beta$ the two Cartesian directions $x, y$. While thermal translational noises are subdominant in active systems, we have nevertheless included them to ensure that thermal equilibrium is achieved in the steady-state with $V_s=0$.
	
	The steric pair potentials between active particles and between active particles and the inclusions, $V({\mathbf r}_{ij})$, and the steric potential between active particles and the channel walls, $V{^\pm_W}(y_i)$ (with $\pm$ indicating the top/bottom walls), are modeled using the modified forms of the Weeks-Chandler-Andersen potential (WCA) 
	\begin{equation}
		\!\!\!\!V({\mathbf r_{ij}}) \!=\!
		\left\{\begin{array}{l l}
			\!\!4\epsilon\! \left [\! \left ( \frac{\sigma_{\mathrm{eff}}}{|{\mathbf r_{ij}}|} \right )^{12}\! \!-\!2\! \left ( \frac{\sigma_{\mathrm{eff}}}{|{\mathbf r_{ij}}|} \right )^{6}\!\!+\!1\right ] &: |{\mathbf r_{ij}}| \!\leq\! \sigma_{\mathrm{eff}}, \\ 
			\!\!0 &: |{\mathbf r_{ij}}| \!>\! \sigma_{\mathrm{eff}},
		\end{array}\right.
	\end{equation}
	\begin{equation}
		\!\!\!V{^\pm_W}(y_i) \!=\!
		\left\{\begin{array}{l l}
			\!\!4\epsilon\! \left [\! \left ( \frac{a}{|{\delta y^\pm_i}|} \right )^{12}\!\!\! -\!2\! \left ( \frac{a}{|{\delta y^\pm_i}|} \right )^{6}\!\!+\!1\right ] 
			&:  |{\delta y^\pm_i}|\! \leq \!a, \\ 
			\!\!0 
			&:  |{\delta y^\pm_i}| \!> \!a,
		\end{array}\right.
	\end{equation}
	where $\epsilon$ will be fixed at a sufficiently large value to ensure that particles, the inclusions and the channel walls remain impermeable. We use $\sigma_{\mathrm{eff}}=2a$ for the interaction of the $i$th and the $j$th active particles, in which case 	$|{\mathbf r_{ij}}|$ represents the center-to-center distance between the active particles. For the interaction of an active particle and an inclusion, we use $\sigma_{\mathrm{eff}}=a+a_c$ and $|{\mathbf r_{ij}}|$ is replaced with the corresponding center-to-center distance. Also, $|\delta y^\pm_i| = |y_i \mp H/2|$ gives the perpendicular distances of the \emph{i}th active particle from the top/bottom wall. Needless to say that the inclusions here are placed in a manner that no steric forces are applied to them except by active particles.

%
	In order to proceed further, we use a dimensionless representation by rescaling the units of length and time using the radius of active particles and the timescale for their translational diffusion as $\tilde x = {x}/{a}$, $ \tilde y = {y}/{a}$ and $ \tilde t = {D_T t}/{a^{2}}$. The relevant set of dimensionless parameters that determine the overall behavior of the system are thus obtained as follows: The size ratio $a_c/a$, the area fraction $\phi$, the rescaled intersurface distance $\tilde{\Delta} = \Delta/a$ (see Fig. \ref{fig:Sketch}), the {\em rescaled chirality strength} $\Gamma=\omega/D_R$ and the {\em P\'eclet number} (or rescaled self-propulsion strength), $Pe_s$, defined as 
	\begin{equation}
		Pe_s=\frac{a V_s}{D_T} = \frac{3 V_s}{4D_R a},  
	\end{equation}
	where we have used $D_R=3D_T/4a^2$ for no-slip spheres in the low-Reynolds-number (Stokes) regime \cite{happel:book1983}. Also, the typical radius of curvature, ${R}_{\omega}$, for the circular arcs traversed by chiral bath particles	will be relevant in the next section; in rescaled units, $\tilde{R}_{\omega}={R}_{\omega}/a $, we have
	\begin{equation}
		\tilde{R}_{\omega} = \frac{4P_{s}}{3\Gamma}.  
	\end{equation}
	The force components on the $i$th active particle and the interaction energy are rescaled as $\tilde f^x_i = -\partial \tilde U/\partial \tilde x_i$, $\tilde f^y_i = -\partial \tilde U/\partial \tilde y_i$ and $\tilde U=U/(k_{\mathrm{B}}T)$, respectively.

	Equations (\ref{Eq:langevin_a}) and (\ref{Eq:langevin_b}) are solved numerically using Brownian Dynamics methods by discretizing them over sufficiently small timesteps $\Delta \tilde t$. 	Typical simulations were run for $10^7$ timesteps using $\Delta \tilde t\sim 10^{-5}$, with $5\times10^6$ initial steps used for relaxation. Then results are averaged over about 20 statistically independent samples after the system reached a steady state.

	Our main objective is to study the role of self-propulsion, intrinsic chirality of active particles, and also the channel height on the effective forces and torques induced on the inclusions. Accordingly, we fix $\epsilon/(k_{\textup{B}}T)=100$, $a_c/a = 5$ and $\phi = 0.3$ and vary $Pe_s$ in the range $0-150$, $\Gamma$ in the range $0$ to $40$, and for $H$ we use two values of $2.5a_c$ and $4.5a_c$. We use 1600 active particles with initial positions distributed randomly in the free space within the simulation box. The latter is defined by the channel height in $y$ direction and a periodic side length in $x$ direction, whose value is adjusted according to the given area fraction.

\begin{figure*}
	\begin{center}
		\includegraphics[width=0.9\textwidth]{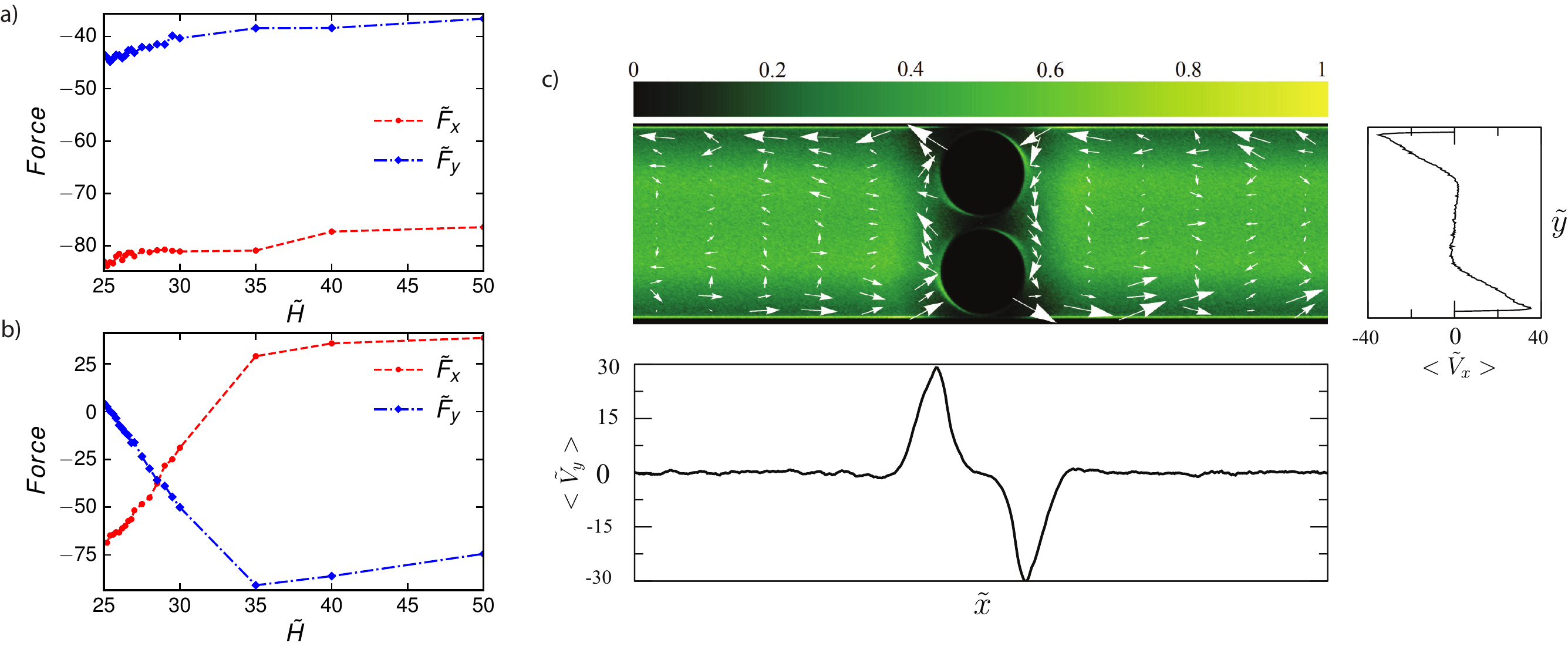}
		\caption{\textbf{a}: The $x$- and $y$- components of effective  force, $\tilde{F}_{x}$ and $\tilde{F}_{y}$, respectively, on the inclusions in the parallel configuration as a function of rescaled channel height, $\tilde{H}$ at fixed $Pe_s = 50$ and $\phi = 0.1$. \textbf{b}: Same as \textbf{a} but in the perpendicular configuration. \textbf{c}: Local area fraction of active bath particles (as shown by color map) along with the averaged velocity of bath particles indicated by arrows. The right and bottom panels show averaged $\tilde{V}_x$ and $\tilde{V}_y$ components of the bath particles velocity, respectively.
		}
		\label{fig:fixed_0p1}
	\end{center}
\end{figure*}

	
	For the calculation of the effective forces on the inclusions, we considered two separate cases. In the first case, the inclusion pair were fixed in place with parallel or perpendicular configuration relative to the midline of the channel ($x$-axis). In this case, the net force acting on a inclusion follows from the averaged sum of instantaneous forces exerted on it from collisions by bath particles (see Refs \cite{Zaeifi:JCP2017,Zarif:PRE2020}). The components of force $\tilde{F}$ are reported by $\tilde{F}_x$ and $\tilde{F}_y$ in $x$ and $y$ directions, respectively. For concreteness, in the forthcoming plots and in the parallel (perpendicular) configuration, we include only the effective force on the right (top) inclusion. Thus, a positive (negative) force indicates a repulsive (attractive) interaction between the inclusions. When we consider a freely rotating inclusion dimer (Section \ref{sec:mobile}), the total torque $\tilde{\boldsymbol{\tau}}$ on the dimer is calculated using the definition $\tilde{\boldsymbol{\tau}} = \tilde{{\mathbf r}}\times  \tilde{{\mathbf F}} = \tilde{\tau} \hat{\mathbf z}$, where $\tilde{{\mathbf r}}$ is the distance vector of the said inclusion from the center-of-mass of the dimer and $\hat{\mathbf z}$ is the out-of-plane unit vector. 

	\section{Results}
	\label{sec:results}  
	
	\subsection{Fixed inclusions}
	\label{sec:fixed}

	We begin our discussion by exploring the impact of particle chirality and channel confinement on the effective forces imparted on the inclusions in both their parallel and perpendicular configurations relative to the channel walls. The results can be compared with those of the nonchiral case in Ref. \cite{Zarif:PRE2020}. Simulation data for the rescaled effective force components, $\tilde{F_{x}}$ and $\tilde{F_{y}}$, in the parallel configuration are shown in Fig.~\ref{fig:fixed_0p1}a as functions of the rescaled channel height, $\tilde{H}$, at fixed values of $Pe_s$ = 50, $\phi = 0.1$, $\Gamma= 19.2$ and $\tilde{\Delta} = 2.0$. As seen, $\tilde{F_{x}}$ is negative and indicates attractive interaction in $x$-direction. By increasing the channel height, absolute values of the force decrease. Positive $\tilde{F_{y}}$ values indicate an `upward' force is experienced by the right-side inclusion (and a `downward' force of equal magnitude by the left-side inclusion), implying a net torque on the dimer, which we will explore further in Section~\ref{sec:mobile}.

	In the perpendicular configuration, see Fig.~\ref{fig:fixed_0p1}b, $\tilde{F_{x}}$, which will now be indicative of a torque on the dimer, changes its sign and magnitude as the channel height, $\tilde{H}$, is increased. $\tilde{F_{y}}$ vanishes at $\tilde{H} = 25$, indicating no force on the inclusions in $y$-direction; the reason being that the distance between the inclusions and the walls are less than the diameter of an active particle. Also the circular motion of active particles and their radius of arc for this P\'eclet number and chirality are such that the accumulation near walls is low. By increasing the rescaled channel height, active-particles are able to go between walls and inclusion, thus a decrease in $\tilde{F_{y}}$ (increase in the attractive force {\em magnitude}) can be seen. On further increase of the height, $\tilde{F_{y}}$ decline to a minimum at $\tilde{H} = 35$, where the attractive force is maximum, after that by increasing $\tilde{H}$, absolute values of force decline while still remaining attractive. In summary, $\tilde{F_{y}}$ acts as an attractive force, while $\tilde{F_{x}}$ changes sign that indicate the torque will be reversed upon increasing the channel height; see Section \ref{sec:mobile}.

	Mean local area fraction of active-particles and their speed profiles in $x$ direction, $\tilde{V}_x$, and in $y$ direction, $\tilde{V}_y$, in the perpendicular configuration for $\tilde{H} = 25$, $Pe_s = 50$, $\phi = 0.1$, and $\Gamma = 19.2$ are displayed in Figure \ref{fig:fixed_0p1}c. This figure shows that most of the active particles, stay away from the walls and inclusions. This observation seems rather peculiar and in contrast to the nonactive or nonchiral active cases where active particles aggregate near edges and inclusions. The velocity profiles indicate a net current inside the channel induced by the collective behavior of active particles, which will be explored in more detail in the case of mobile colloidal inclusions in Section~\ref{sec:mobile}.

	Comparing Figs.~\ref{fig:fixed_0p1}a and \ref{fig:fixed_0p1}b, it should be noted that the distance between inclusions and the walls of the channel is not the same. For the case of parallel configuration of inclusions, (Fig.~\ref{fig:fixed_0p1}a), due to the high distance between the  inclusions and the wall, the intersections between active-particle ring around the inclusions and a flat active-particle layer on the walls are absent compared to the perpendicular configuration (Fig.~\ref{fig:fixed_0p1}b). The population of active particles in the layer and ring decreases as we move away from the pertinent surfaces. Therefore, in the parallel case, intersections between low populated layer and ring of the inclusions create a weaker force compared to intersections due to relatively high-populated layer and ring when inclusions are located perpendicular to the channel walls. 

\begin{figure*}
	\begin{center}
		\includegraphics[width=0.9\textwidth]{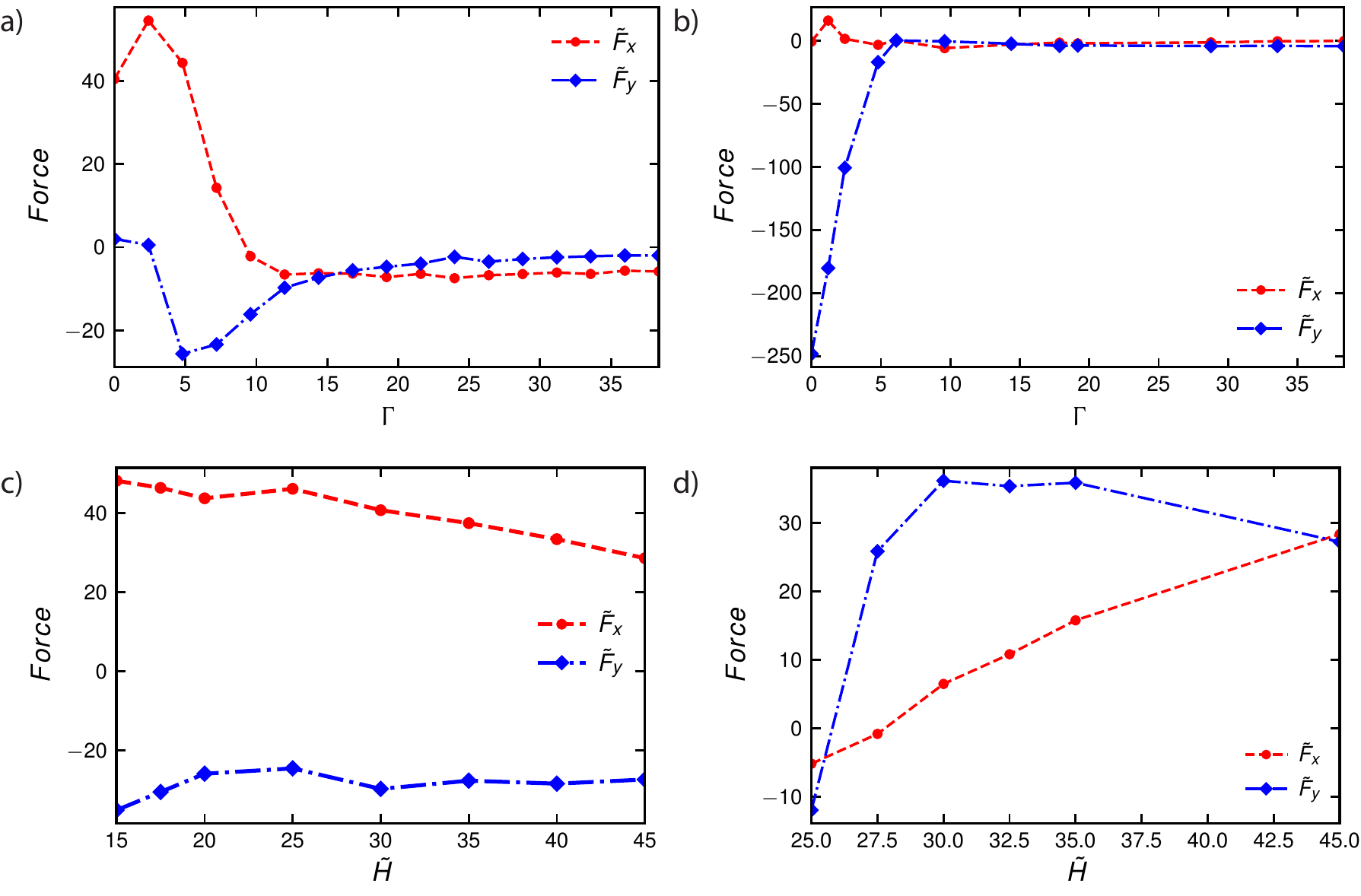}
		\caption{The effective  force components ($\tilde{F_x}$ and $\tilde{F_y}$)  on the inclusions as a function of $\Gamma$ for channel height $\tilde{H} = 25$ and $Pe_s = 50$ and in \textbf{a}: parallel and \textbf{b}: perpendicular configurations. The effective force components ($\tilde{F_x}$ and $\tilde{F_y}$) on the inclusions as a function of $\tilde{H}$ in \textbf{c}: parallel and \textbf{d} perpendicular configurations at $\Gamma = 4.8$ and $Pe_s = 50$.
		}
		\label{fig:fixed_0p3}
	\end{center}
\end{figure*}
	We now turn to the case of a moderately larger area fraction $\phi = 0.3$. First, we consider the effect of chirality on the force. Figures \ref{fig:fixed_0p3}a and \ref{fig:fixed_0p3}b show forces acting on the inclusions at different chiralities, $\Gamma$, at fixed height, $\tilde{H} = 25$, and P\'eclet number $Pe_s = 50$, for parallel and perpendicular configurations, respectively. In the parallel case, with $\Gamma = 0$ (no chirality), $\tilde{F_y} = 0$ for the right-side inclusion indicating no overall torque, also $\tilde{F_{x}}$, the force acting on the right-side inclusion is positive, meaning mediated repulsive forces between the two inclusions, as expected from the nonchiral active cases. However, by increasing chirality, forces acting on the inclusions reach a maximum at around $\Gamma = 2.4$ (corresponding to $\tilde{R}_{\omega} \sim \tilde{H}$), then decline to reach small values after $\Gamma = 9.6$ (corresponds to $\tilde{R}_{\omega} \sim \tilde{H}/4$). It can be seen that $\tilde{F_{y}}$ is negative, which implies a counterclockwise torque on the inclusions. The said values of the circular arc radius for the bath particles, respectively, represent a situation, where a full rotation of bath particles (in the absence of rotational diffusion) can happen inside the channel, and a situation, where, roughly, two circular trajectories can fit vertically into the channel. In the first case, the inclusions experience the maximum force from the chiral active bath, making the inclusions rotate in the same direction as bath particles, however, in the second case, building two circular collective motions, one close to the top wall and the other close to the bottom wall, cancels out each other at the center of the channel and for the  inclusions creates a weak opposite rotation of the bath particles.

	In the perpendicular case, at $\Gamma = 0$, $F_y$ values are high, due to an increased build-up of active particles density at the regions between the inclusions and walls while $\tilde{F_x}$ values are minimal. For the nonchiral case ($\Gamma = 0$), the number of flat layers formed at the walls from the accumulation of particles is high and these layers can overlap with the ring(s) formed around the inclusions leading to an attractive force. However, as expected \cite{Zaeifi:JCP2017}, the activity-induced attractive force is suppressed upon increasing the chirality strength, $\Gamma$. As it is known, in the limit of $\Gamma \rightarrow \infty$, the effective force mediated on inclusions from active-chiral bath particles approaches its equilibrium values \cite{jamali2018active}.

	Data for forces acting on the inclusions as a function of channel height at $\Gamma = 4.8$ and $Pe_s = 50$ are shown in Figs.~\ref{fig:fixed_0p3}c and \ref{fig:fixed_0p3}d for parallel and perpendicular configurations of the inclusions, respectively. The difference in the force behavior can be understood in terms of ring-layer interactions. In fact, for both figures (\ref{fig:fixed_0p3}c and \ref{fig:fixed_0p3}d), magnitude of the rescaled radius of curvature for the bath particles is the same, $\tilde{R}_{\omega}/a = 4P_{s}/(3\Gamma) \sim 13.5$. The mean area fraction occupied by the swimmer particles can be used to demonstrate the number of layers (rings) formed for walls (inclusions). The number of layers and rings are not affected by the channel height \cite{Zaeifi:JCP2017,Zarif:PRE2020}, and it is only affected by the area fraction of the system, $\phi$, chirality, and P\'eclet number. In the case of confined nonchiral active particles, the effective  force is highly affected by channel height and due to the overlaps between layers of flat boundaries with the rings produced around the inclusions, creates an oscillating force profile. In the channel heights with high overlaps, the force magnitude increases and otherwise shows a decrease in force. However, for the case of chiral active particles, the number of layers and rings diminished and as a result, similar to other cases, the oscillating behavior has been suppressed. In the parallel case, a suppressed oscillating behavior for the force can be seen, which it will reach the bulk value at wider channels. For perpendicular configuration, on the other hand, starting at $\tilde{H} = 25$, $\tilde{F_{y}}$ values are negative, meaning attractive forces between the inclusions, while $\tilde{F_{x}}$ values indicate a counterclockwise torque. This can be understood from layer-ring and ring-ring overlap competitions, which at high confinement layer-ring overlaps dominate ring-ring overlaps leading to attractive interactions between the two inclusions. However, as the channel gets wider, layers-rings overlap weakens and ring-ring overlaps dominate, creating repulsive force. Finally, the repulsive force reaches a maximum at $\tilde{H}\sim 35$, then decreases until reaching a plateau at the wider channel.

\begin{figure}
	\begin{center}
		\includegraphics[width=3.0in]{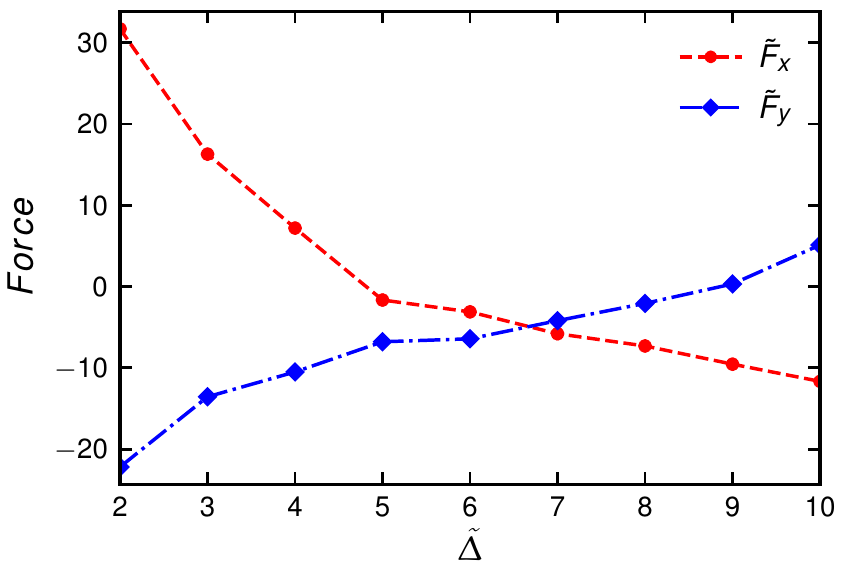}
		\caption{Force components as functions of the distance, $\tilde{\Delta}$ , on inclusions in the parallel configuration at $\Gamma = 4.8$, $Pe_s$ = 50 and $\tilde{H} = 15$.
		}
		\label{fig:Fixed_distance_0p3}
	\end{center}
\end{figure}

	We also studied the effect of distance between the inclusions on the mediated force at $\Gamma = 4.8$ and $Pe_s$ = 50 in parallel configuration and the result is depicted in Fig.~\ref{fig:Fixed_distance_0p3}. The $\tilde{F_{x}}$ values decrease with increasing distance, which is to be expected as ring-ring overlaps weaken when the distance between the inclusions is increased. $\tilde{F_{y}}$ values starting from $\tilde{\Delta} = 2$ are negative, implying clockwise torque acting on the inclusions (similar rotation as active particles), increasing $\tilde{\Delta}$ decreases absolute values of the force. As the inclusions are separated further, the mediated force gradually becomes independent of the presence of the other inclusion and approaches to a single-inclusion case.

\subsection{Rotating inclusion dimer}
\label{sec:mobile}
	In this section, we explore the effect of mediated force on inclusions by allowing them to rotate freely as a rigid dimer, while their center of mass is pinned to the center of the channel. Here, we will obtain further insights on the effects of chirality and active-particle interactions with the walls and inclusions.

\begin{figure}
	\begin{center}
		\includegraphics[width=3.0in]{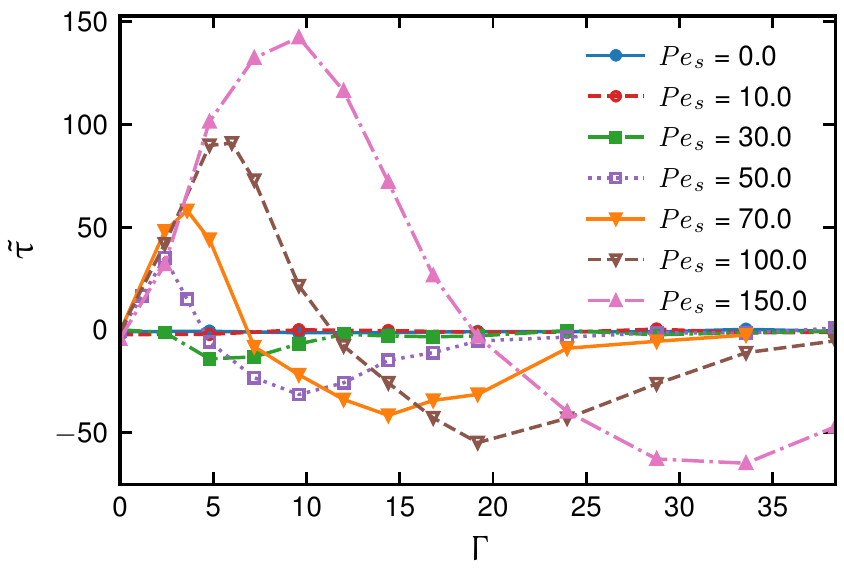}
		\caption{Torque, $\tilde{\tau}$, acting on the inclusion dimer as a function of $\Gamma$ and $Pe_s$ at fixed channel height $\tilde{H} = 25$.}
		\label{fig:torque_all_Pe}
	\end{center}
\end{figure}

	We start by calculating average torque, $\tilde{\tau}$, acting on inclusion dimer as a function of $\Gamma$ and $Pe_s$ at fixed channel height $\tilde{H} = 25$. Simulation data, in this case, are illustrated in Fig.~\ref{fig:torque_all_Pe}. There are some main features in this figure that needs exploring in more detail. It is evident that with increasing swimmer $Pe_s$ the absolute values of torque acting on dimer increases. As expected, in nonactive particles ($Pe_s = 0$) there is no net torque acting on the dimer as a result of the symmetry of the system. Also, as mentioned earlier, positive(negative) torque means counterclockwise (clockwise) rotation of dimer. A maximum and a minimum are observed for the torque, corresponding to the maximum speed of counterclockwise and clockwise rotation of the dimer, respectively. The maximum and minimum values of torque happens at lower values of chirality as the P\'eclet number decreases, such that, the maximum for the cases $Pe_s = 10$ and $30$ are not seen from our simulations.
	
	At a given P\'eclet number, by introducing the chirality to the active particles, their dynamics change from active to active chiral motion, in which, as the value of chirality increases the radius of curvature decreases, $\tilde{R}_{\omega} = V_s/\omega$. As a consequence of changes in the motion, the spatial distribution of active particles changes inside the channel, leading to a decrease in the population of particles at boundaries and an increase in the population of them away from boundaries. For chiral active particles the most important parameter, determining their behavior, is the radius of curvature. For specific values of this radius which allows the active particles to have a complete rotation inside the channel, maximum values of forces are acting on inclusions. However, owing to the density of the system, layers of active particles are present in the proximity of walls, causing the radius of arc for a perfect rotation to decrease from the channel height to smaller values. Also, the value of the P\'eclet number determines the persistence time of active particles close to the boundaries. At higher values of the P\'eclet number, it takes longer for the swimmer particles to escape from the boundaries. As a result, for the case of $Pe_s = 50$, the torque corresponding to a perfect rotation inside the channel has its maximum value where $\Gamma \sim 2.4$ (or $\tilde{R}_{\omega} \sim 27.16$), and its minimum value at $\Gamma \sim 9.6$ (or $\tilde{R}_{\omega} \sim 6.9$). Similar behavior can be seen at different P\'eclet numbers. Table~\ref{tab:PeOmegamaxmin} summarizes the values corresponding to the maximum and minimum of the torque depicted in Fig.~\ref{fig:torque_all_Pe}. The origin of this behavior was initially shown and discussed in the force measurement in the previous section. Depending on the layers and rings around the inclusions and walls, and their interactions, the force acting on the  inclusions is different, which in free mode causes the dimer to rotate in the applied direction.

		\begin{table}
			\begin{centering}
				\begin{tabular}{|p{1.5cm}|p{1.5cm}|p{1.5cm}|p{1.5cm}|p{1.5cm}| }
			\hline
			$Pe_s$ & $\Gamma_{\rm{max}}$ & $\tilde{R}_{\omega}$ & $\Gamma_{\rm{min}}$ & $\tilde{R}_{\omega}$ \\ \hline
			50   & $2.4$ & $\sim 27.16$ & $9.60$ & $\sim 6.79$ \\ \hline
			70   & $3.6$ & $\sim 25.35$ & $14.4$ & $\sim 6.34$ \\ \hline
			100  & $6.0$ & $\sim 21.73$ & $19.2$ & $\sim 6.79$ \\ \hline
			150  & $9.6$ & $\sim 20.38$ & $33.6$ & $\sim 5.82$ \\ \hline
				\end{tabular}
			\par\end{centering}
			\centering{}\caption{\label{tab:PeOmegamaxmin} Values of the radius of arc for maximum and minimum torques at given P\'eclet numbers.
			}
		\end{table}



\begin{figure}
		\begin{center}
			\includegraphics[width=3.5in]{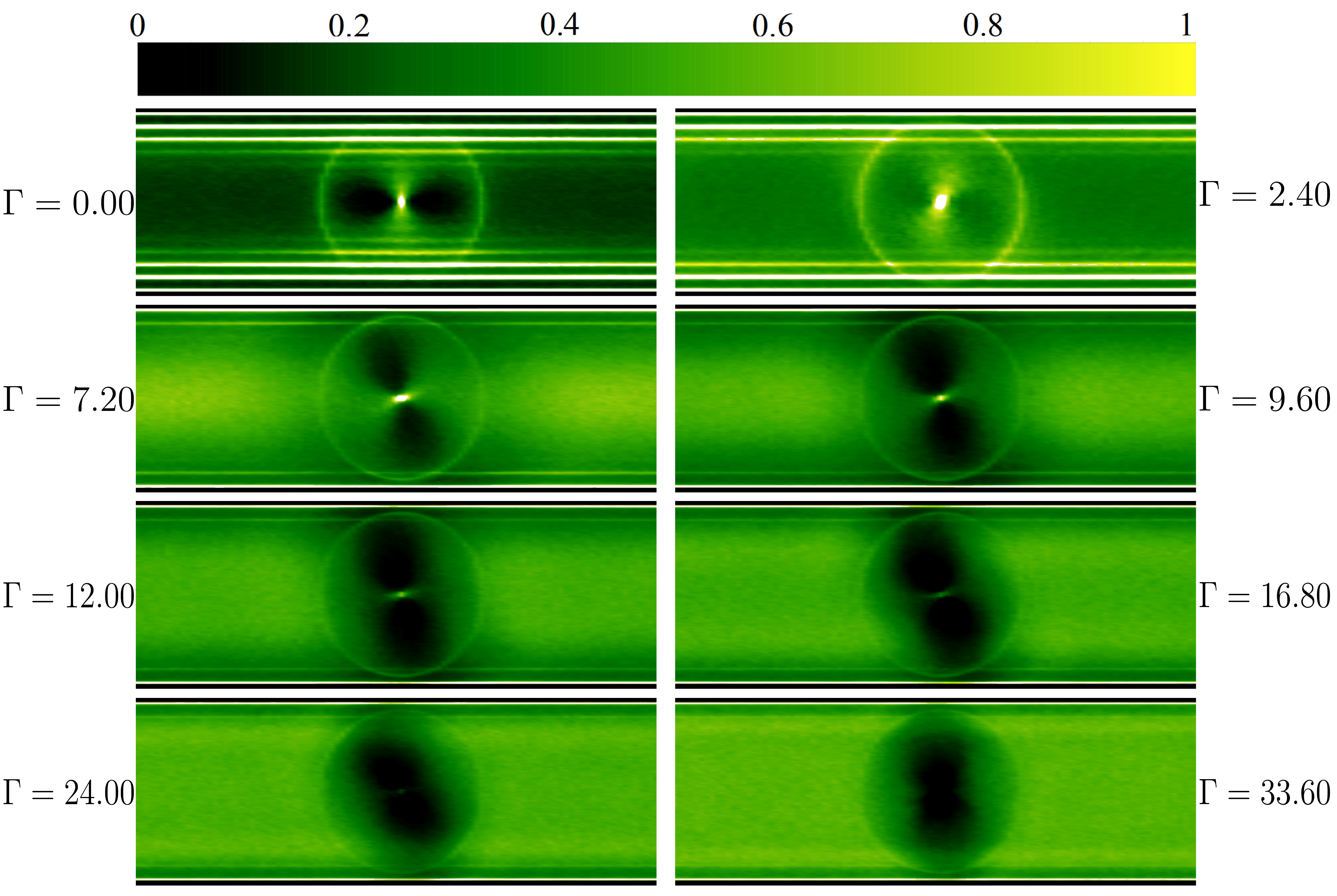}
			\caption{Density map of active-particles for various $\Gamma$s, as indicated in the figure, at fixed values of $\phi$ = 0.3, $\tilde{H} = 25$ and $Pe_s$ = 50.
			}
			\label{fig:density_profile}
		\end{center}
	\end{figure}

	Density map of active-particles for various chiralities ($\Gamma$) at fixed values of $\phi$ = 0.3, $\tilde{H} = 25$ and $Pe_s = 50$ are shown in Fig.~\ref{fig:density_profile}. Staring at $\Gamma = 0$, about 5 layers of active particles reside in the vicinity of the walls. Also, active-particle density where dimer spends most of its time is minimum (black regions). The density of active particles is high in the center of the channel because active particles can go between the inclusions. By increasing $\Gamma$, fewer layers are formed near the walls, and the density of active particles in the middle of the channel is increased. Increasing chirality from $\Gamma=9.6$ to $\Gamma=12$ leads most of the active particles to the middle of the channel and only two layers remain near the walls. On increasing $\Gamma$ further, the number of layers near the walls decreases, and more active particles move to the middle of the channel. 

\begin{figure}
	\begin{center}
		\includegraphics[width=3.0in]{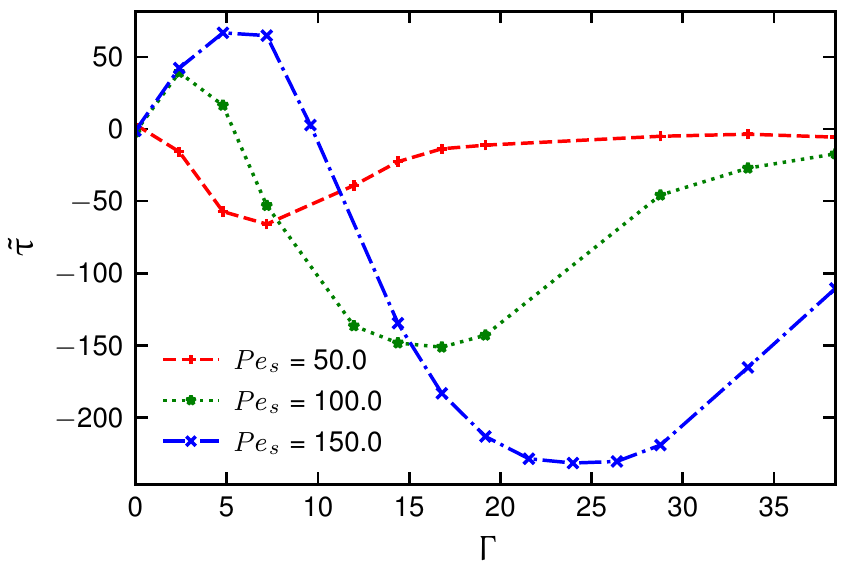}
		\caption{Torque, $\tilde{\tau}$, acting on inclusion dimer as a function of chirality $\Gamma$ for swimmer $Pe_s$ = 50, 100 and 150 with fixed $\phi$ = 0.3 and channel height $\tilde{H} = 45$.
		}
		\label{fig:Torque_Height_Pe}
	\end{center}
\end{figure}

	To determine the effects of channel height, the values of torque, $\tilde{\tau}$, acting on dimer as a function of chirality ($\Gamma$) of active particles (we only considered cases with $Pe_s$ = 50, 100 and 150) at fixed $\phi = 0.3$ and channel height $\tilde{H} = 45$ were calculated and shown in Fig.~\ref{fig:Torque_Height_Pe}. Compared to the results reported in Fig.~\ref{fig:torque_all_Pe}, the position of maxima and minima in the torque acting on inclusions from bath particles is moved to lower chiralities (as a result, in the case of $Pe_s = 50$, the maximum is not seen at the simulated chiralities). The values of arc radius for the swimmer particles in both figures (Figs.~\ref{fig:torque_all_Pe} and \ref{fig:Torque_Height_Pe}) are the same, however, the available space for active particles to rotate has increased in the wider channel, resulting in a change in the relative population of active particles along the channel. Therefore, using a similar argument as $\tilde{H} = 25$, the maximum of the torque occurs as the radius of the arc is close to the channel height and the minimum occurs as $\tilde{R}_{\omega} \sim \frac{\tilde{H}}{4}$.

\begin{figure}
	\begin{center}
		\includegraphics[width=3.0in]{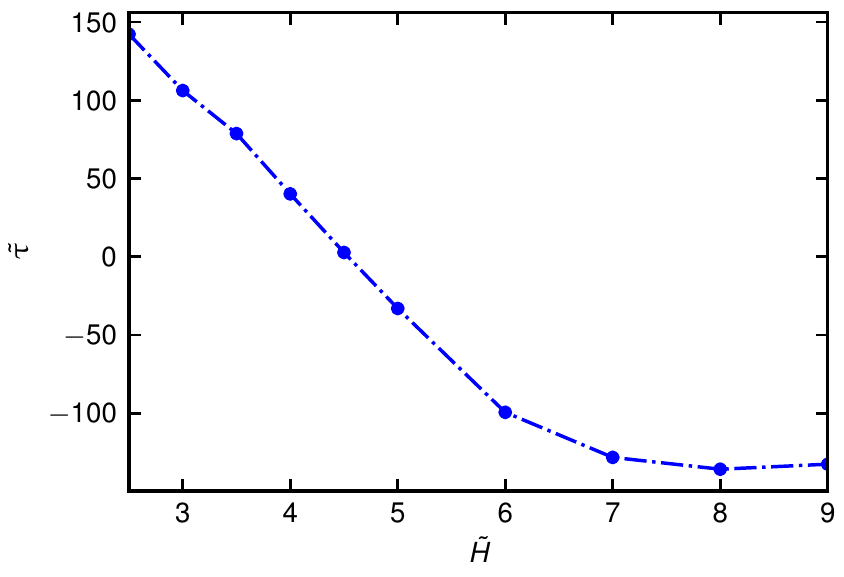}
		\caption{Torque, $\tilde{\tau}$, acting on inclusion dimer as a function of channel height $\tilde{H}$ with fixed chirality, $\Gamma$ = $12$, $\phi$ = 0.3, and swimmer P\'eclet number, $Pe_s$ = 150. Last point in the plot refers to torque in bulk situation.
		}
		\label{fig:torque_H_Pe_150}
	\end{center}
\end{figure}

	We also calculated the torque, $\tilde{\tau}$, acting on inclusion dimer as a function of channel height, $\tilde{H}$, with fixed chirality, $\Gamma$ = $12$, $\phi$ = 0.3, and swimmer P\'eclet number, $Pe_s$ = 150, as depicted in Fig.~\ref{fig:torque_H_Pe_150}. By increasing $\tilde{H}$; torque value which was the maximum value for $\tilde{H}=25$ decreases and passes zero torque at $\tilde{H} \sim 45$ and decreasing more until reaching its minimum value at $\tilde{H}=80.0$. After that torque increases slowly tending to a fixed value of bulk for this particular $\Gamma$. Comparing the torque values applied to pinned dimer from bath particles, we reach a similar conclusion as we did for the cases of constant channel heights at $\tilde{H} = 25$ and 45. At this particular P\'eclet number and chirality, the value for the radius of arc is $\sim 16.3$, meaning a maximum torque occurs at $\tilde{H} \sim 2 \tilde{R}_{\omega}$ and minimum at $\tilde{H} \sim 4 \tilde{R}_{\omega}$.

\begin{figure}
		\begin{center}
			\includegraphics[width=3.0in]{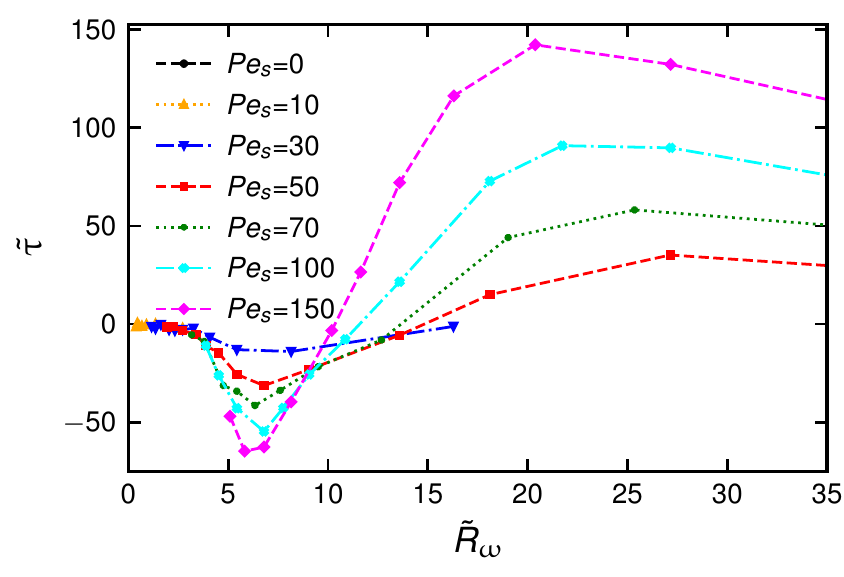}
			\caption{Data of torque acting on the inclusion dimer as plotted in Fig.~\ref{fig:torque_all_Pe}. Here, instead of chirality, torque values are plotted as a function of circular arc radius of the active-particles calculated for all swimmer $Pe_s$.
			}
			\label{fig:Torque_vs_Radius}
		\end{center}
	\end{figure}

	As a conclusion, here we present the Simulation data for torque as a function of circular arc radius of the active particles calculated for all swimmer $Pe_s$ in Fig.~\ref{fig:Torque_vs_Radius} (the data is the same as in Fig.~\ref{fig:torque_all_Pe}, however re-plotted as a function of $\tilde{R}_{\omega}$). The absolute values of average torque acting on dimer increase with swimmer $Pe_s$, its maximum and minimum values appear in specific values of radius for all $Pe_s$, namely at $\sim 7$ and $\sim 23$, respectively. These values correspond to $\sim \frac{\tilde{H}}{4}$ and $\sim \tilde{H}$ as we have discussed in previous sections.

	\section{Conclusion}
	\label{sec:Conclusion}
	The effect of chirality and confinement on the behavior of colloidal inclusions in an active bath of particles is studied using Brownian dynamics simulations. The goal of the current work was to describe how the motion of active particles is affected by applying confinement on the system and observe the changes this can cause on the mediated force on the inclusions. In particular, we considered how the collective motion of chiral active particles emerges in the presence of confinement and studied this behavior by measuring mediated force on the pinned inclusion dimer. 
	
	Our study of the force mediated by active-chiral bath particles, reveals the importance of the spatial distribution of active particles on the force applied to the inclusions. The spatial distribution of bath particles inside the channel is determined mainly by the radius of the circular motion of particles. Two colloidal particles located inside a nonactive medium experience predominately attractive force between them, known as depletion interaction. In the presence of planner walls, they are also attracted by the walls. However, in the case of the active medium, these forces become nonmonotonic with more determining factors, such as the magnitude of confinement, activity of bath particles, chirality, the area fraction of bath particles, aspect ratio of the inclusions/bath particles, etc. In the active confinement case, the interactions are repulsive, due to the high persistence time of active particles at the boundaries. Thus, the active particles accumulate at surfaces to create layers and rings for the walls and the inclusions, respectively. Therefore, the interaction between the inclusions is determined by the interactions of layers and rings \cite{Zaeifi:JCP2017,Zarif:PRE2020}. 
	
	In the case of chiral active bath, chirality suppresses the effect of activity on the force mediated on the inclusions \cite{Zaeifi:JCP2017}. Chirality decreases the persistence time of active particles at boundaries. This makes layers and rings less populated and causes a decrease in the interaction's magnitude. At elevated chiralities, the spatial distribution of bath particles changes dramatically, compared to the nonchiral case, as a result, the mediated interactions between the inclusions become attractive again. In addition, our results show that the inclusion configuration inside the channel plays a major role as well.
	
	To quantify the effect of chirality on the inclusions, we measured the torque applied from bath particles to pinned inclusion dimer. The result in our work shows a nonmonotonic behavior for the torque as a function of the chirality of the bath particles. For a given self-propulsion and channel height, the torque shows a maximum and a minimum at the studied range of chiralities. The maximum values of torque are found to be connected to the arc radius equal to the channel height, while, the minima are related to the arc radius equal to one-quarter of the channel height, where two complete rotations can fit in the channel.
	
	An oscillating behavior is expected to happen for the torque as the channel gets wider, creating room for more circular motions to locate inside the channel. Therefore, the channel height can be used as a control parameter for a given chiral-active system to mediate the desired rotation on the inclusions. This can be used as an advanced technique to mix fluid in a selective and directive way.

	For a given channel width it is interesting to explore the effect of colloid/swimmer aspect ratio on the behavior of the dimer. As this ratio increases, it is expected to see an increase in the magnitude of the torque, however, the occurrence and position of the maximum and minimum values need to be confirmed. Another interesting direction can be studying the effect of shape asymmetry of the inclusion (such as the snowman model \cite{Milinkovi:PRE2013}) in the presence of chiral active particles. 
	
	\section{Acknowledgements}
	\label{sec:Acknowledgments}
	We acknowledge computational resources provided by the Center for High-Performance Computing (SARMAD) at Shahid Beheshti University, Tehran.
		
	\bibliography{ref}
	
\end{document}